\documentclass[twocolumn,superscriptaddress]{revtex4-1}
\usepackage{amsmath,amsthm,amssymb} \usepackage{hyperref,graphicx}
\usepackage{color}
\usepackage{natbib}

\begin{document}

\title{Kinetics of motility-induced phase separation and swim pressure}

\author{Adam Patch} \affiliation{Department of Physics and Soft Matter Program, Syracuse University,
Syracuse, NY 13244} 

\author{David Yllanes} \affiliation{Department of Physics and Soft Matter Program, Syracuse University,
Syracuse, NY 13244}  \affiliation{Instituto de Biocomputaci\'on y F\'isica de
Sistemas Complejos (BIFI), 50009 Zaragoza, Spain}

\author{M. Cristina Marchetti}\affiliation{Department of Physics and Soft Matter Program, Syracuse University, Syracuse, NY 13244}  

\date{\today}

\newcommand{\david}[1]{{\bfseries \color{green} #1}}
\newcommand{\adam}[1]{{\bfseries \color{blue} #1}}

\begin{abstract} Active Brownian particles (ABPs) represent a minimal model of active matter consisting of self-propelled spheres with purely repulsive interactions and rotational noise.  Here, we examine the pressure of ABPs in two dimensions in both closed boxes and systems with periodic boundary conditions and show that its nonmonotonic behavior with density is a general property of ABPs and is not the result of finite-size effects. We correlate the time evolution of the mean pressure towards its steady state value with the kinetics of motility-induced phase separation. For parameter values corresponding to phase separated steady states, we identify two dynamical regimes. The pressure grows monotonically in time during the initial regime of rapid cluster formation, overshooting  its steady state value and then quickly relaxing to it, and remains constant during the subsequent slower period of cluster coalescence and coarsening. The overshoot is a distinctive feature of active systems.  
\end{abstract}

\maketitle \section{Introduction} 

Over the past decade, there has been growing interest in the physics of active 
matter, defined as collections of self-driven particles that exhibit rich emergent 
behavior~\cite{Marchetti2013}. Realizations abound in the living world, from bird 
flocks~\cite{Vicsek1995,Toner1995,Ballerini2008a} to 
epithelial cell monolayers~\cite{Szabo2006,Saez2005}, 
and in engineered systems, such as self-catalytic colloids~\cite{Paxton2004,Ginot2015} 
and microswimmers \cite{Elgeti2015}. The distinctive property of active systems 
is that the particles are independently driven out of equilibrium and time reversal 
symmetry is broken locally, rather than globally as in systems driven out of 
equilibrium by external fields or boundary forces.

Progress has recently been made in formulating the nonequilibrium statistical
mechanics of active matter using a minimal model of active Brownian particles
(ABPs) consisting of purely repulsive self-propelled spherical colloids with
overdamped dynamics~\cite{Fily2012a}. Perhaps the most remarkable property of
this simple system is that it spontaneously phase separates into a dense liquid
phase and a gas phase in the absence of any attractive
interactions~\cite{Tailleur2008,Fily2012a,Fily2014,Redner2013,Cates2015,Marchetti2016a}.
This phenomenon arises from the persistent dynamics of self-propelled particles
when the time for particles to reorient after a collision exceeds the mean free
time between collisions, hence the name of motility-induced phase separation
(MIPS). Additionally, in a sort of reverse MIPS,  confined ABPs
spontaneously accumulate at the 
walls of the container~\cite{Elgeti2013,Yang2014,Fily2015},
a behavior which is at odds with  fundamental properties of  gases and fluids
in thermal equilibrium.  These findings have raised a lot of interest  in
understanding whether active matter can be characterized in terms of
equilibrium-like properties, such as effective temperature and pressure.
A broad class of active particles can exert persistent forces on the walls of a container. These forces have been  quantified recently in terms of a new contribution 
to the pressure, dubbed swim pressure, that measures the flux of propulsive forces 
across a unit bulk plane of material~\cite{Yang2014,Takatori2014}. Remarkably, it 
has been shown that in generic active systems the mechanical force exerted 
on the walls of the container depends on the detail of particle-walls
interactions, making it impossible to define the pressure of an active fluid as
a state function~\cite{Solon2015}.

This result, however, does not apply to the special case of spherical ABPs. For
this minimal model it has been shown that the pressure is indeed a state
function that characterizes the bulk materials properties of this simple active
fluid \cite{Solon2015, Takatori2015b}. On the other hand, the behavior
of pressure of spherical ABPs  is unusual, in that simulations have reported a
non-monotonic behavior of pressure versus density, arising from the suppression
of motility (and therefore of swim pressure) associated with particle caging at
the onset of phase separation~\cite{Yang2014,Winkler2015}. Experimental
measurements of pressure of active colloids have similarly shown a strong
suppression of pressure at intermediate density~\cite{Ginot2015}.
In spite of extensive work, some open question remains concerning the
quantitative  role of finite-size effects in active systems and the origin and
robustness of the non-monotonic behavior of pressure.

In this paper we use molecular dynamics simulations to investigate the 
pressure of spherical ABPs with soft repulsive forces and correlate  its
non-monotonic behavior with density with the kinetics of cluster size growth in
phase separating systems.  By examining both systems bounded by confining walls
and ones with periodic boundary conditions, we show that finite-size effects
are consistent in both cases with the behavior expected for a rarefied thermal
gas. The pressure calculated in confined systems is strongly suppressed by the
presence of boundaries if the persistence length of the particles dynamics is
comparable to the linear size of the container, with a linear dependence on
system size and a behavior that resembles that of a rarified Knudsen gas. With
periodic boundary conditions the convergence to the large system size limit is
exponential, again as expected in a thermal system. 

The non-monotonic behavior of
pressure as a function of density for repulsive ABPs with large persistence
lengths was first reported in simulations of systems in closed boxes~\cite{Yang2014} and has been seen in sedimentation experiments of active colloids~\cite{Ginot2015}. More recently  it was  confirmed in numerical studies of hard active particles 
 with periodic boundary conditions in three dimensions~\cite{Winkler2015}.
Our work provides a systematic study of finite size effects in both bounded and periodic systems and shows that the nonmonotonicity is not a finite size effect, but a bulk properties of ABPs. It arises becasue in phase separted systems the aggregate effectively provides a bounding wall for the active gas, which suppresses the swim pressure. Additionally, by considering soft repulsive disks, we complement previous work on hard particles and show that this effect does not depend on the details of the interparticle interaction.  Finally,  we examine
the kinetics of coarsening and establish a strong correlation between the aggregation dynamics and the relaxation of the 
pressure towards its steady state value.  We identify two dynamical regimes that control the relaxation of a
disordered initial state towards the nonequilibrium steady state of the system.
For system parameters that produce a homogeneous steady state,
pressure grows monotonically in time to its steady state value,
which it reaches on a timescale controlled 
by the persistence time of the self-propelled dynamics.
In contrast, for parameters that produce a phase-separated steady state,
the time evolution of the pressure is not monotonic and there are two 
dynamical regimes. Initially, small clusters form and break up rapidly
and the pressure quickly builds up and overshoots its asymptotic value.
Eventually, the dynamics crosses over to a coarsening regime with a slower cluster growth
and the pressure relaxes to its steady-state value.

The rest of the paper is organized as follows. Section II described the ABP
model and provides details of our simulations. Section III examines the swim
pressure and its behavior at low and high density. It also  discusses the
relevance of finite-size effects.  Section IV examines correlations between the
pressure relaxation and the kinetics of MIPS of swim pressure. Finally, we close in Section V with some concluding remarks.

\section{Active Brownian Particles Model}

We consider a well-established minimal model of monodispersed ABPs in two dimensions~\cite{Fily2012a}, consisting of $N$ self-propelled disks of radius $a$ in a square box of area $L^2$. Each particle is identified by the position $\boldsymbol{r}_i$ of its center and a unit vector $ \hat{\boldsymbol{e}}_i = ( \cos\theta_i , \sin\theta_i )$ that defines the axis of self  propulsion. Assuming the medium only provides friction, the dynamics is governed by Langevin equations, given by
\begin{align} 
  \dot{\boldsymbol{r}}_i &= v_0 \hat{\boldsymbol{e}}_i + \mu \sum_{j\neq i} \boldsymbol{F}_{ij} \;, \\
  \dot\theta_i &= \eta_i(t)\;,
\end{align} 
where $v_0$ is the bare self-propulsion speed, directed
along $\hat{\boldsymbol{e}}_i$ and  $\eta_i(t)$
a Gaussian random torque with zero mean and variance
$\langle\eta_i (t)\eta_j (t')\rangle = 2D_\text{r}\delta_{ij}\delta(t-t')$, with $D_\text{r}$
the rate of rotational diffusion. The pair forces $\boldsymbol{F}_{ij}$ are purely repulsive and harmonic, with $\boldsymbol{F}_{ij} = k ( 2 a - r_{ij} )
\hat{\boldsymbol{r}}_{ij}$ for $r < 2a$ and $F_{ij} = 0$ otherwise, where $\boldsymbol{r}_{ij}=\boldsymbol{r}_i-\boldsymbol{r}_j$ is the interparticle separation and $\hat{\boldsymbol{r}}_{ij}=\boldsymbol{r}_{ij}/|\boldsymbol{r}_{ij}|$. Finally, $\mu$ is the mobility. 
We neglect here noise in the translational dynamics, which is less important than the orientational one in both synthetic active colloids and swimming bacteria~\cite{Marchetti2016a}. The nonequilibrium nature of this minimal active model is provided entirely by the propulsive force $\boldsymbol{F}^\text{s}_i=(v_0/\mu)\hat{\boldsymbol{e}}_i$. After integrating out the angular dynamics, $\boldsymbol{F}_{i}^\text{s}$ represents a non-Markovian stochastic force  correlated over the persistence time $\tau_\text{r}= D_\text{r}^{-1}$. Since the finite correlation time of the noisy propulsive force is not matched by similar correlations in the friction  coefficient $\mu^{-1}$, which is constant, the system does not obey the equilibrium fluctuation-dissipation theorem embodied by the Stokes-Einstein relation. 

\begin{figure} 
\includegraphics[height=0.5\textwidth,angle=270]{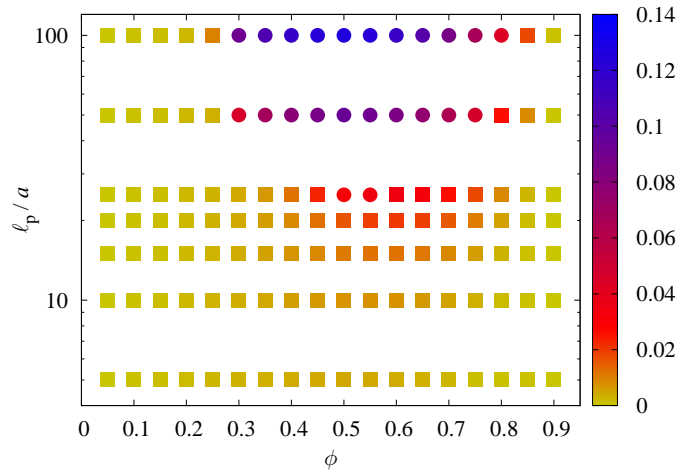} 
\caption{Phase diagram of ABPs for a system of size $L=200$ in the plane of packing fraction 
$\phi$ and persistence length $\ell_\text{p}$. The phase diagram is constructed by examining the probability distribution $P(\phi_\text{W})$ of the local densities, computed by dividing the system in $N_\text{W}$ windows of size $20\times20$ and calculating  the packing fraction $\phi_\text{W}$ in each window. The heat map shows the values of the variance of this distribution,  defined as
$\sigma^2_\text{W} = \frac{1}{N_\text{W}} \sum_{i=1}^{N_\text{W}}\langle( \phi_\text{W}-\phi)^2\rangle$.
In  phase-separated states, the probability 
distribution of $\phi_\text{W}$ is bimodal (see, e.g., \cite{Redner2013}) and  $\sigma_\text{W}^2$ is large. We show with circles those values of the parameters that result in a $P(\phi_\text{W})$ with two peaks (gas density and aggregate density).
\label{fig:phase_diagram}} 
\end{figure}

The  persistence time $\tau_\text{r}$ controls the crossover from  ballistic to
diffusive regime in the single particle dynamics. For $t\gg\tau_r$ the dynamics of noninteracting ABPs is diffusive, with diffusion coefficient $D_\text{s}=v_0^2\tau_r/2$. The single particle dynamics can also be characterized by the persistence length, 
$\ell_\text{p} = \frac{v_0}{D_\text{r}}$, and the rotational P\'eclet 
number, $\text{Pe}_\text{r}=\ell_\text{p}/a$, used in much of the literature~\cite{Redner2013,Marchetti2016a}. 

Our simulations employ a conventional Brownian dynamics
algorithm~\cite{Branka1999}. 
We take the interaction timescale, $\tau_\text{D} = (\mu k)^{-1}$, 
as the unit time ($\mu=k=1$) and the particle radius $a$ as the unit of length ($a=1$). 
To prevent particles from passing through each other, we set $v_0 = (a \mu
k)/100$. We choose a time step sufficiently small to handle many-body
interactions at high density, $\delta t \ll \tau_\text{D}$.

As discussed in the introduction, purely repulsive ABPs exhibit macroscopic phase separation, or MIPS, where the dense phase grows to the size of the system and nearly free particles in the gas phase 
try to explore lengths or order $\ell_\text{p}$.  For this reason, ABPs are subject to strong finite-size effects that we quantify here by considering  simulation boxes of linear size  $L=50,100,200,400$, both in a closed box geometry
and with periodic boundary conditions. We tune the packing fraction $\phi = N \frac{\pi a^2} {L^2}$ that sets 
the total number of particles, $N$, with $N \sim 10^5$ for $L=400$. A schematic representation of the region of the phase diagram explored in the present work is shown in Fig.~\ref{fig:phase_diagram}, where the color represents the variance of the local density. Phase-separated systems are plotted with circles.

All simulations have been run for a  time $t_\text{f}$ of $10^6$ time units or longer, which in all cases is several orders of magnitude
longer than the time scale of the particles' rotational diffusion: $t_\text{f}\gtrsim 500 D_\text{r}^{-1}$. To examine the
time evolution of our observables, we average the relevant physical quantities
over exponentially increasing time windows. In particular, we use the so-called
$\log_2$-binning procedure by following the evolution of averages in time
intervals $I_n = ( 2^{-(n+1)} t_\text{f},\ 2^{-n} t_\text{f}]$. The error bars 
are estimated from the fluctuations between several independent runs with
a jackknife procedure (see, e.g., \cite{Amit2005}). We have $500$ 
runs  for $\phi=0.01$, $10$--$20$ runs for larger values of $\phi$.

We also use this time binning procedure to assess convergence to the steady state.
We consider the simulation to have reached a steady state if the averages in at least 
the first three $I_n$ are compatible with each other within errors.  If this condition 
is not met, we double the simulation time. In particular for parameter values where 
the system phase separates, simulations have been run for $10^7$ time units.

Finally, we shall denote our estimate of the steady state ensemble average of an 
observable $O$ by $\langle O\rangle$, which we compute by taking the value for the 
$I_0$ time interval (i.e., the last half of the simulations).

\begin{figure}[t]
\includegraphics[height=0.49\textwidth,angle=270]{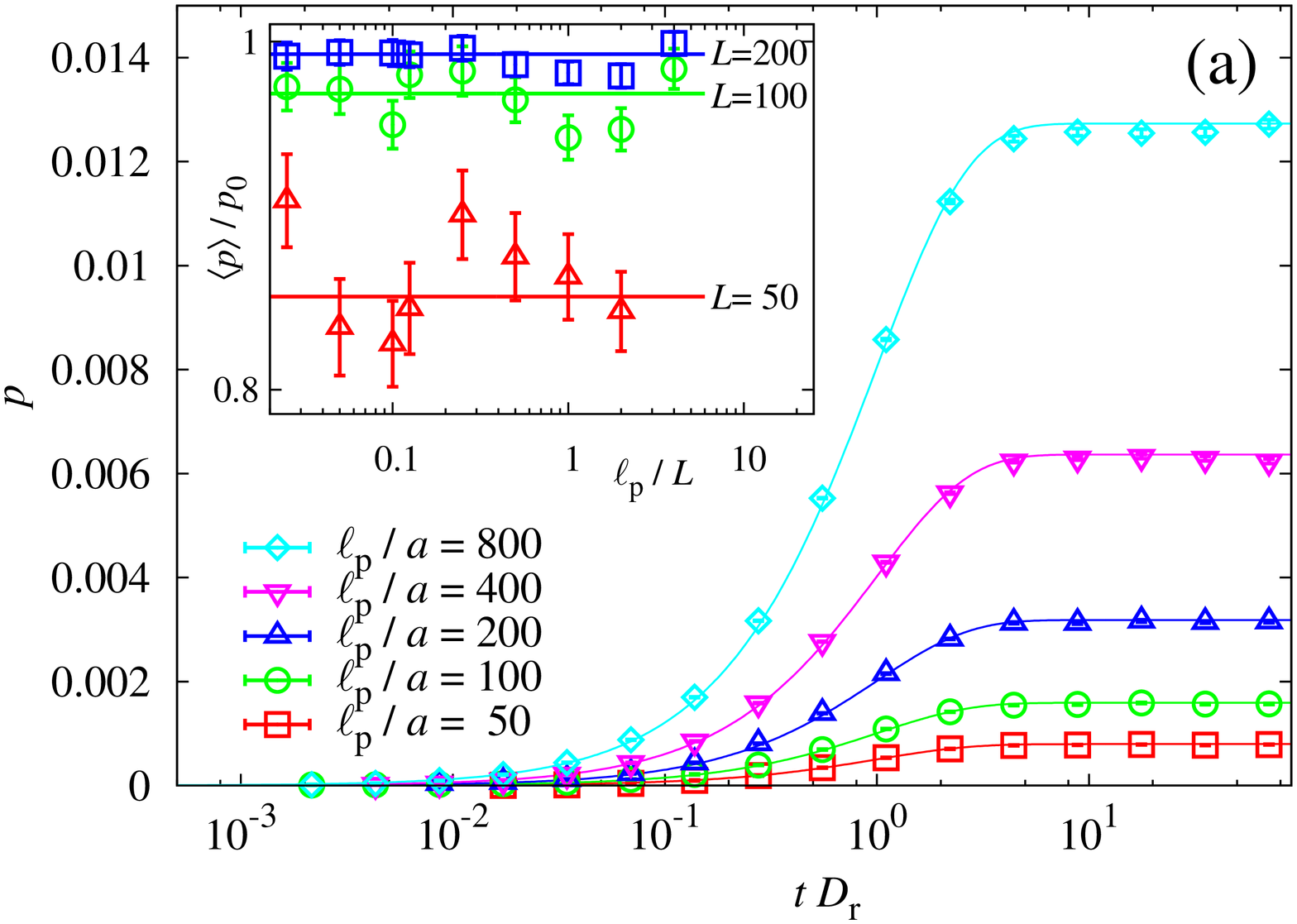} 
\includegraphics[height=0.49\textwidth,angle=270]{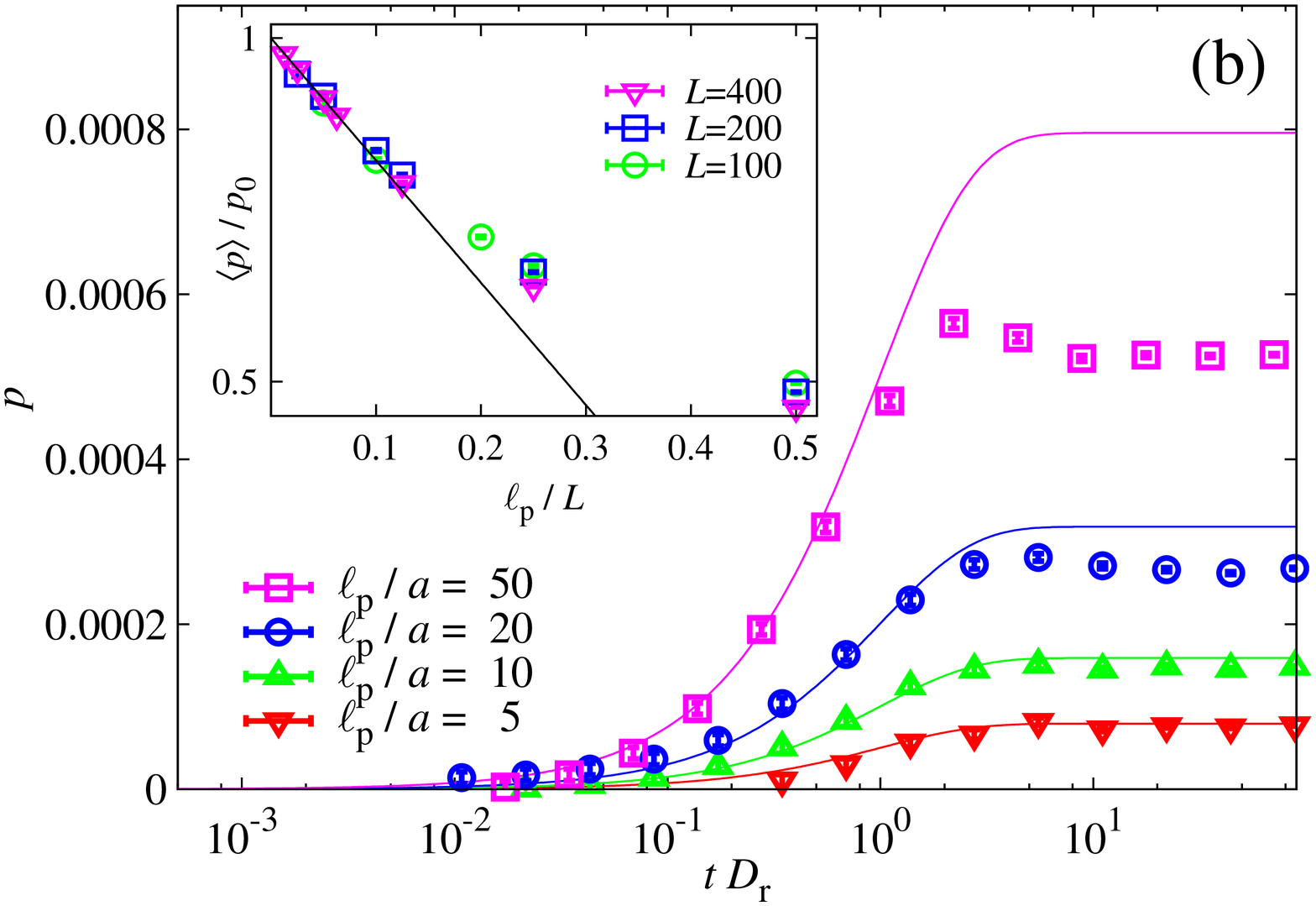}
\caption{Time evolution of the pressure of a dilute gas ($\phi=0.01$) of ABPs  
for various values of the persistence length for periodic (a) and closed (b) 
square boxes of linear size $L=200$. At this density the pressure is determined entirely 
by the swim pressure. The solid lines show the large-$L$ ideal gas prediction of 
Eq.~\eqref{eq:p_vs_t}, which matches the numerical results for periodic boundary 
conditions even at very large $\ell_\text{p}$. For closed systems the pressure is 
significantly suppressed by accumulation of particles at the walls as soon as 
$\ell_\text{p}$ becomes comparable to $L$.
The insets show our estimate for the steady-state pressure $\langle p \rangle$ in the dilute limit divided by the ideal gas steady state value $p_0 = \rho v_0 \ell_\text{p} / 2 \mu$
as a function of persistence length for a various systems sizes $L$. 
For periodic boundary conditions (a)-inset the system is self-averaging and 
$\langle p\rangle^{(L)}\approx p_0$ within errors for $L=200$. 
For the closed box (b)-inset, there are strong finite-size effects due
to the walls. We show that $\langle p\rangle/p_0$ versus $\ell_\text{p}/L$
can be fit to $1-A\ell_\text{p}/L$ for $\ell_\text{p} / L \leq 0.1$ (black line)
with a result of $A=1.78(3)$, $\chi^2=7.16$ for $6$ degrees of freedom (d.o.f.).
\label{fig:dilute_p}}
 \end{figure}

\section{Swim pressure}

There has recently been a lot of interest in characterizing the pressure of active systems. Pressure can be defined in terms of the forces transmitted across a unit bulk plane of material. In an active fluid there is a unique contribution to pressure that measures the flux of propulsive forces across a bulk plane of material~\cite{Yang2014,Takatori2015b,Mallory2014}. This contribution, named swim pressure in recent literature~\cite{Takatori2014c}, can be expressed via a virial-type formula as 
~\cite{Yang2014,Takatori2014,Marchetti2016a}
\begin{align}
\label{eq:p_s} 
p_\text{s} = \frac{1}{dL^2} \sum\limits_{i} \boldsymbol{F}^\text{s}_i \cdot
\boldsymbol{R}_i \;, 
\end{align} 
with $d$ the spatial dimension (here $d=2$). 
In a closed box $\boldsymbol{R}_i=\boldsymbol r_i$, but with periodic boundary
conditions $\boldsymbol R_i$ is  the position of the particle in an infinite
system, accounting for winding numbers as the particle crosses the periodic
boundary~\cite{Louwerse2006,Winkler2009}. The virial expression estimates the swim pressure as the propulsive force carried over a distance of the order of the persistence length, in analogy to the kinetic pressure of an ideal gas or the radiation pressure of a photon gas.  Note, however, that in spite of the one-body expression given in Eq.~\eqref{eq:p_s}, the swim pressure depends on interactions that lead to suppression of the persistence length.
In addition to the swim pressure, there is also the contribution describing the direct transmission of interaction forces across the bulk plane. For pairwise forces $\boldsymbol{F}_{ij}$, as relevant to our collection of ABPs, this can be calculated from the familiar virial expression
\begin{equation} 
\label{eq:p_D} 
p_\text{D}= \frac{1}{dL^2} \sum\limits_{i,j} \boldsymbol{F}_{ij}
\cdot \boldsymbol{r}_{ij} \;.
\end{equation} 
Recent work has shown that for the minimal model of spherical ABPs considered here the total pressure $p = p_\text{s}+p_\text{D}$ defined from Eqs.~\eqref{eq:p_s} and \eqref{eq:p_D} coincides with the force per unit area on the walls of a container~\cite{Yang2014} and  represents a state function of the active fluid~\cite{Solon2015}, independent of the properties of the walls. 

In the remainder of this section we examine the importance of finite-size effects in the calculation of the pressure of active systems and demonstrate that the predicted~\cite{Yang2014} and observed~\cite{Ginot2015} non-monotonicity of pressure versus density is an intrinsic property of these nonequilibrium fluids, not  a finite-size effect.

\subsection{Pressure of a dilute active gas} 

For a dilute gas of ABPs, the dominant
contribution to the pressure comes from the swim pressure $p_\text{s}$.  Neglecting interactions, and in the large-size limit, 
Eq.~\eqref{eq:p_s} can be calculated exactly, with the result~\cite{Mognetti2013a,Yang2014} 
\begin{align} 
\label{eq:p_vs_t} p(t) &=  p_0 \bigl( 1 - e^{-D_\text{r} t} \bigr)\;, & p_0   
                        &= \rho \frac{ v_0^2}{2 \mu D_\text{r}}\, ,  
\end{align} 
where $\rho=N/L^2=\phi/(\pi a^2)$ is the number density. 
The pressure $p_0$ of an active ideal gas can be naturally interpreted in terms of an 
active temperature, with 
$p_0=\rho k_\text{B}T_{\text{a}}$, 
and  
$T_\text{a} = v_0^2 / ( 2 \mu k_\text{B} D_\text{r})$. This active temperature also 
coincides with the one set by the ratio of translational diffusion $D_\text{s}=v_0^2\tau_r/2$ 
to the  friction coefficient $\mu^{-1}$, as required for thermal Brownian particles 
satisfying the Stokes-Einstein relation. Indeed,  the stochastic 
propulsive force becomes $\delta$-correlated in time in the limit $\tau_r\rightarrow 0$, 
with $k_\text{B}T_\text{a}={\rm constant}$~\cite{Fily2012a}. In other words, noninteracting ABPs behave 
like thermal colloid at temperature $T_\text{a}$ in the limit of vanishing persistence time, $\tau_\text{r}$. 
Here we consider the pressure of active particles for finite values of the persistence time, away
from this Brownian limit.

The time evolution of $p_\text{s}$ of a dilute active
gas ($\phi=0.01$) is shown in Fig.~\ref{fig:dilute_p} for a system in a
box of linear size $L=200$ with periodic boundary conditions (a) and one
enclosed by bounding walls (b), for various values of the persistence
length, $\ell_\text{p}$.  Flat repulsive walls are implemented using the same harmonic
forces that describe interparticle interactions.

For periodic boundary conditions, Eq.~\eqref{eq:p_vs_t} provides an excellent
fit to the calculated pressure over the entire time range and for all
$\ell_\text{p}$ considered.   In confined systems, however, a second time scale
comes into play, that is the time $\tau_L=L/v_0$ it takes an ABP to travel
the size of the box. The finite-size corrections to pressure
are negligible only when $\tau_L\gg\tau_\text{r}$, or equivalently $L \gg
\ell_\text{p}$ (here with $L=200$ this only holds for the smallest
$\ell_\text{p}=5$). In all other cases the presence of confining walls
strongly suppresses the asymptotic long-time value of the pressure as compared
to the value $p_0$ given by Eq.~\eqref{eq:p_vs_t}.
\begin{figure}[t]
\includegraphics[height=0.49\textwidth,angle=270]{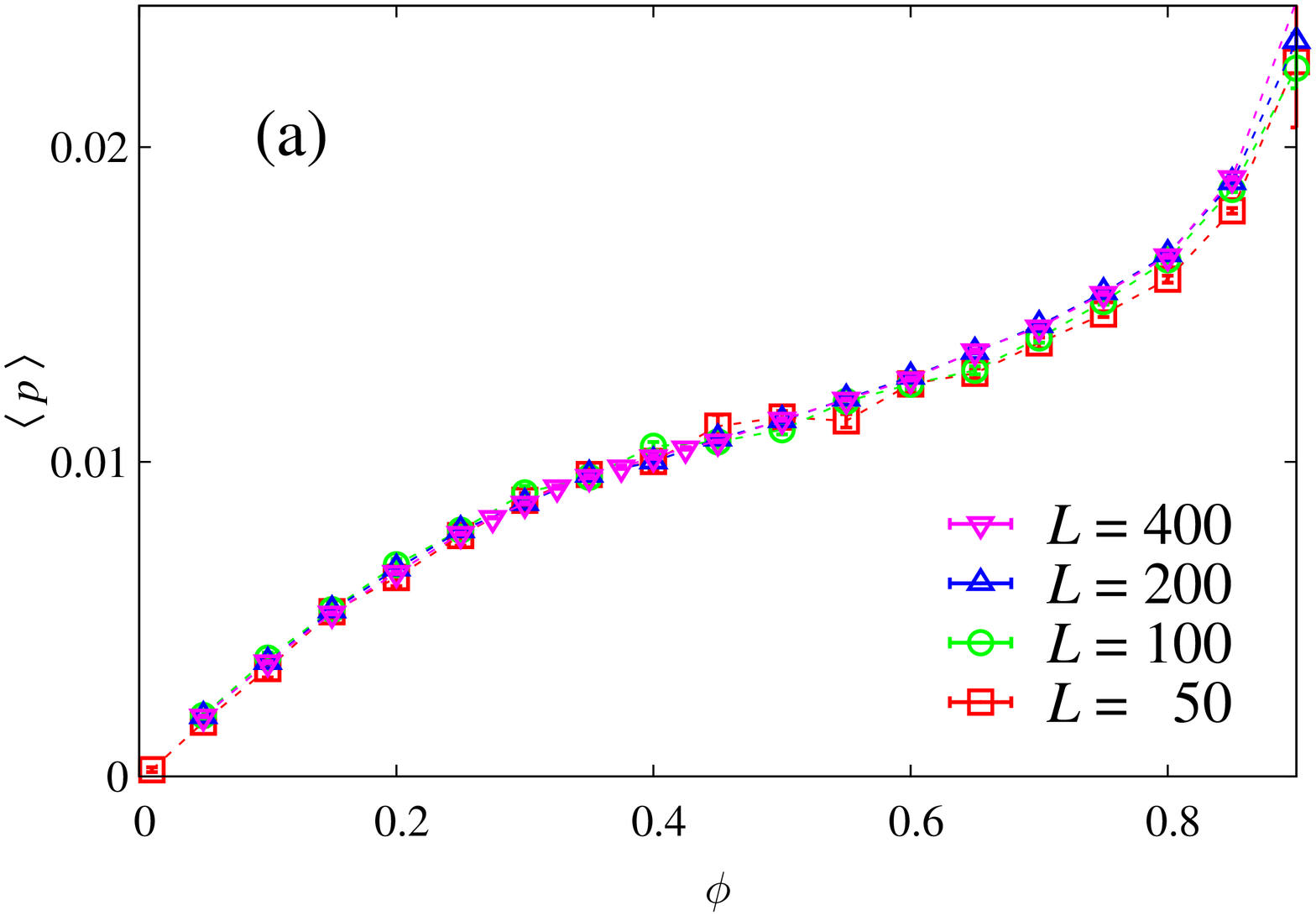}
\includegraphics[height=0.49\textwidth,angle=270]{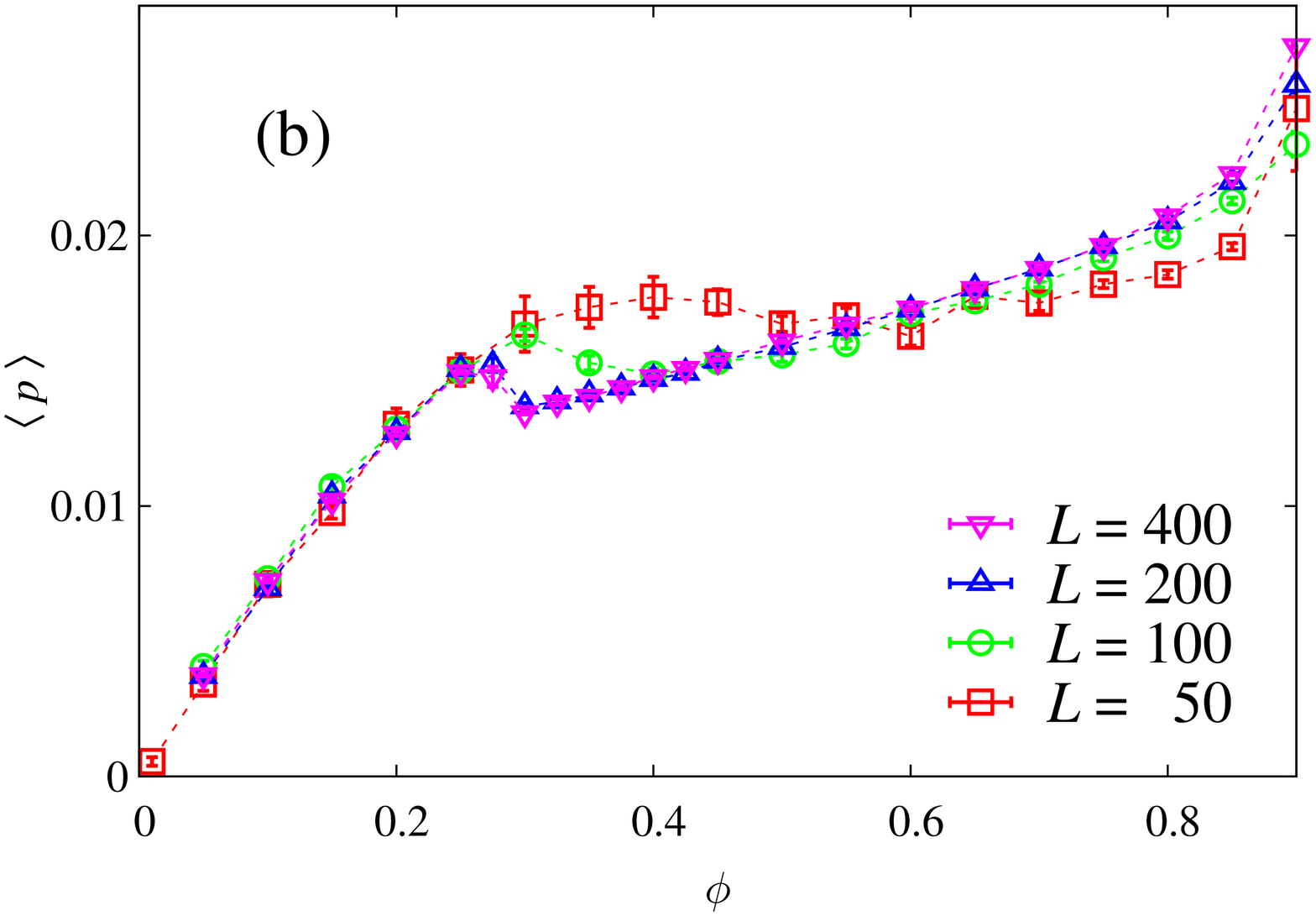}
\caption{Pressure as a function of packing fraction for $\ell_\text{p}/a = 25$ (a) and
$\ell_\text{p}/a=50$ (b) for a system with periodic boundary conditions. The box
size must be several times larger than $\ell_\text{p}$ to obtain results representative
of the large-$L$ limit.\label{fig:p_vs_phi_by_L} }
 \end{figure}
\begin{figure}[t]
\includegraphics[height=0.49\textwidth,angle=270]{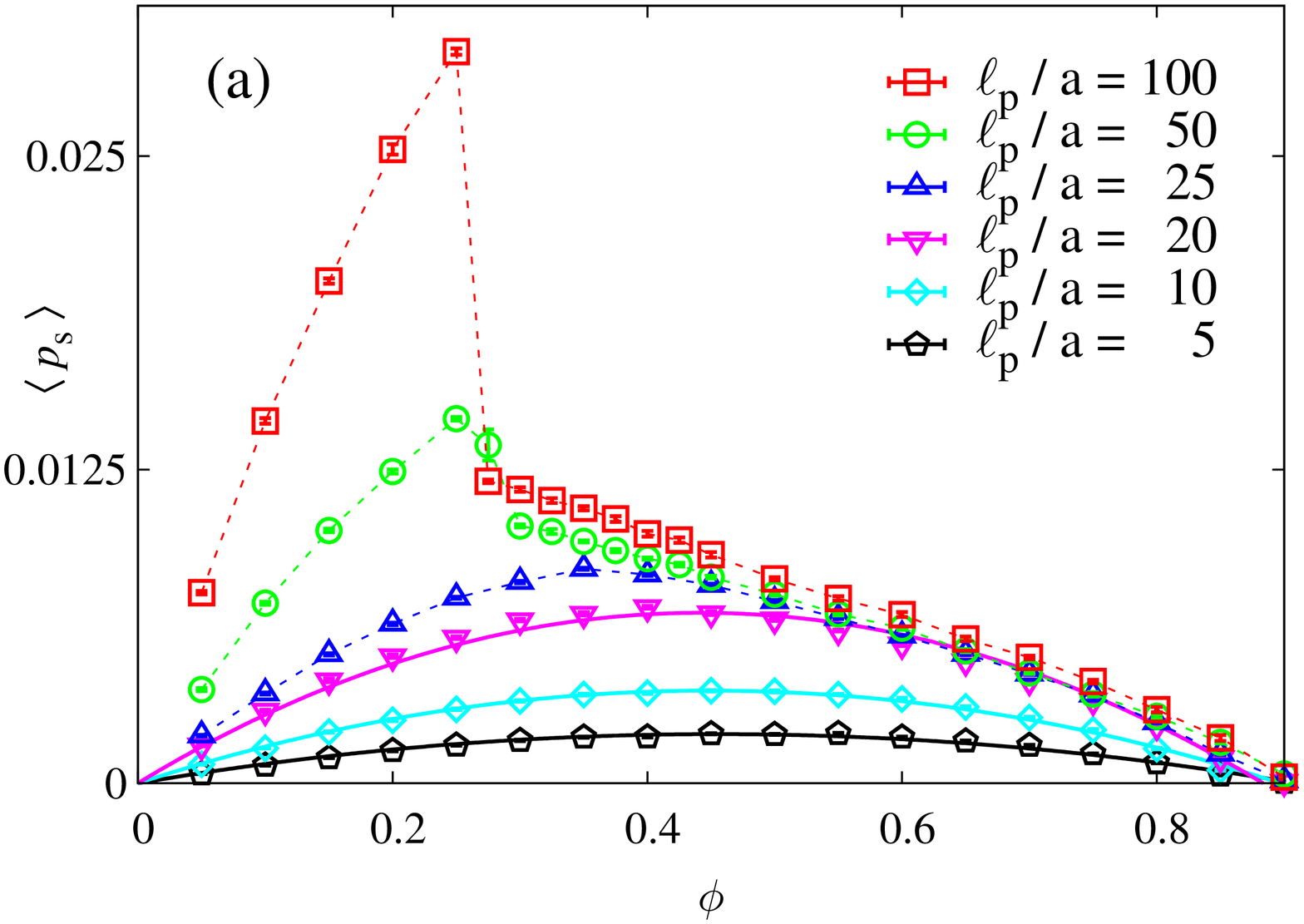}
\includegraphics[height=0.49\textwidth,angle=270]{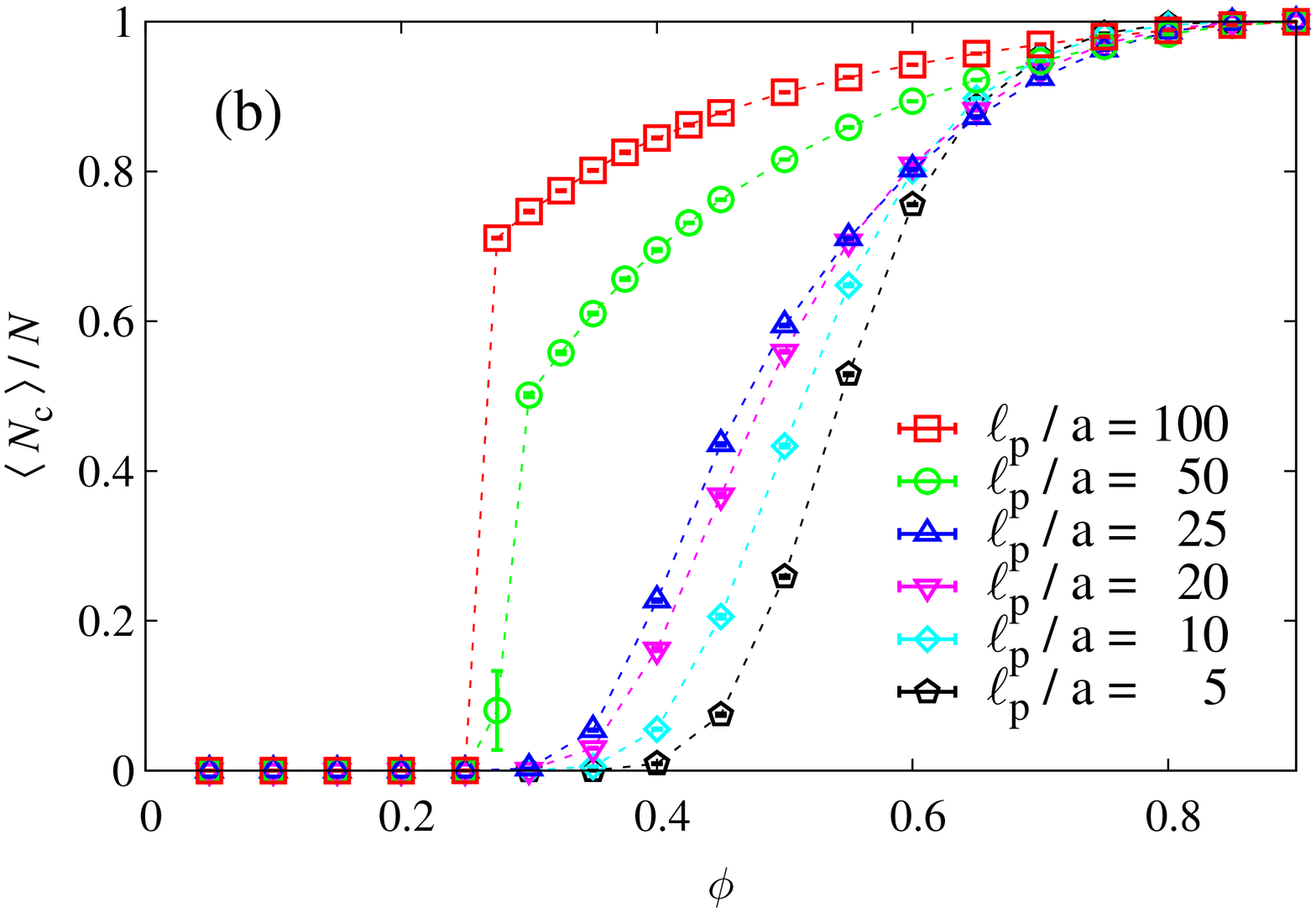}
\caption{Swim pressure $\langle p_s \rangle$  (a) and cluster
fraction $\langle N_\text{c} \rangle/N$ (b) versus packing fraction
$\phi$ for several values of $\ell_\text{p}$ (we use $L=200$ for $\ell_\text{p}/a\leq25$ 
and $L=400$ for $\ell_\text{p}/a=50,100$). 
In (a), solid lines show fits to 
$\langle p_\text{s} \rangle = p_0(\phi) (1-\phi/\phi_0)$, where $\phi_0$ is the 
only adjustable parameter (see Ref.~\cite{Marchetti2016a}). This quadratic behavior breaks 
down for higher $\ell_\text{p}$. In both frames, dashed lines provide a guide for the eye. \label{fig:ps_fc_vs_phi_by_lp}} 
\end{figure}
To quantify the finite-size corrections, the insets of Fig.~\ref{fig:dilute_p} show
the ratio $\langle p\rangle/p_0$ of the calculated
steady-state pressure to the ideal active gas pressure as a function of
$\ell_\text{p}$ for various systems sizes $L$. With periodic boundary conditions
(a), the finite-size corrections are essentially independent of $\ell_\text{p}$ and
converge very quickly to $p_0$ with increasing system size---in fact for
$L=200$ $\langle p(\ell_\text{p})\rangle /p_0$ is indistinguishable
from $1$ with our errors.

In the case of the closed box the finite-size effects are much more pronounced and depend strongly on
$\ell_\text{p}$, as shown in the inset of Fig.~\ref{fig:dilute_p}b. 
Our $\langle p \rangle/ p_0$ for different $L$ can be nearly collapsed when 
plotted versus $\ell_\text{p}/L$. 
In fact, for $\ell_\text{p}/L \lesssim0.1$, the behavior can be 
fitted by a linear form, $\langle p\rangle/p_0 \simeq 1- A \ell_\text{p}/L$. This functional 
dependence supports the idea that the box boundary has a simple surface effect on the pressure 
for $\ell_\text{p}/L \ll 1$. At higher persistence lengths, collective effects, such as 
clustering of particles at the corners of the box,  yield further deviations from the linear behavior.


In summary,  we find that at low density the swim pressure is clearly suppressed in systems that restrict the mean free path below $\ell_p$, so that the dynamics remains ballistic at all times. When particles are confined, there are strong boundary effects reminiscent of those seen in a Knudsen gas, where the density is so low that the mean-free path from inter-particle collisions exceeds the box size. Specifically, the convergence to the large-$L$ limit is exponential for the periodic box and linear for the closed box. This is precisely the type of the finite-size effects one expects in a thermal system whose free energy would have bulk and surface contributions, with the former the only relevant one in periodic systems at large $L$.

\subsection{Pressure at finite density}

At finite density, both the swim contribution and the direct contribution from
interactions are appreciable. The direct
contribution form interactions, $p_\text{D}(\phi)$, grows monotonically with
density as in passive systems and depends only weakly on self propulsion~~\cite{Yang2014}.
In contrast the swim pressure is non-monotonic in density and strongly
suppressed at intermediate density due to the decrease of particle motility.  
This leads to an overall non-convex density dependence of the
total pressure $\langle p(\phi) \rangle$,
which has been seen in simulations~\cite{Yang2014,Solon2015,Winkler2015} and
experiments~\cite{Ginot2015}. For larger $\ell_\text{p}$ simulations
show a non-monotonic $\langle p (\phi)\rangle$.

When the system phase separates, the macroscopic 
aggregate effectively provides a bounding wall to the gas phase, leading
to strong finite-size effects in the swim pressure even
in systems with periodic boundary conditions. This is shown in Fig.~\ref{fig:p_vs_phi_by_L}.
Here, the bottom frame ($\ell_\text{p}/a=50$) shows a system which undergoes phase separation for intermediate densities, while in the top frame ($\ell_\text{p}/a=25$) the phase separation is still incipient (recall Fig.~\ref{fig:phase_diagram}).
We note that in the former case we need box sizes of $L\ge200$ to achieve 
convergence~\footnote{At very high packing fraction, approaching or even surpassing 
the closed-packed limit, there are larger discrepancies, but we do not consider 
this regime here.}.
All the results reported in the remainder of the paper are for 
box sizes $L=200,400$, and are free of finite-size effects within our error bars.

For sufficiently large values of $\ell_p$ the pressure of the active fluid is
non-monotonic with density. This behavior, shown in 
Fig.~\ref{fig:p_vs_phi_by_L}b, is not a finite-size effect and is associated with
the strong suppression of the swim pressure arising from motility-induced
aggregation.  Ref.~\cite{Solon2015} showed that the suppression of
$p_s$ can be captured by a simple expression, given by
\begin{equation} 
  p_\text{s} = \rho \frac{v_0v(\rho)}{2\mu D_\text{r}} \;,
\label{eq:p_quadratic} 
\end{equation}
with $v(\rho)=v_0 + \mu \langle \hat{\boldsymbol{e}}_i
\cdot \sum_{j\neq i} \boldsymbol{F}_{ij}\rangle$  the effective velocity of a particle along its
direction of self-propulsion.
To leading order in the density, this has a linear 
decay $v(\rho) = v_0 (1 - \rho / \rho_* )$, where $\rho_* \equiv \rho_* (\ell_\text{p})$ 
provides a cutoff, above which $v(\rho)$
goes to zero. Inserting this form of $v(\rho)$ into 
Eq.~\eqref{eq:p_quadratic} yields a quadratic form for the swim pressure, 
$p_\text{s}=k_BT_a\rho(1-\rho / \rho_* )$~\cite{Solon2015}.

As we can see in Fig.~\ref{fig:ps_fc_vs_phi_by_lp}a, this
ansatz works well only for moderate $\ell_\text{p}$ ($\ell_\text{p}/a \lesssim25$
with our parameters). For large $\ell_\text{p}$, i.e., for systems exhibiting
phase separation (see Fig.~\ref{fig:phase_diagram}), the swim pressure displays a
much stronger dependence on density  and drops abruptly at the onset of phase 
separation. Phase separating systems always evolve to contain a single large dense region, 
 
The relation between pressure and particle aggregation is shown in Fig.~\ref{fig:ps_fc_vs_phi_by_lp}. The bottom frame displays   
the total number of particles $N_\text{c}$ that belong to clusters above a certain threshold size (we use a cutoff of 100)~\footnote{A particle $i$ is part of cluster $\mathcal{C}$ if it is interacting with any other disks 
in that cluster (i.e., if $r_{ij} \leq 2a$ for any $j \in \mathcal{C}$).}. 
The sharp drop in pressure shown in Fig.~\ref{fig:ps_fc_vs_phi_by_lp}a at large values of $\ell_\text{p}$ corresponds to a jump in the value of $N_\text{c}$, signaling
the onset of phase separation. A similar result has been obtained for hard repulsive spheres in $d=3$~\cite{Winkler2015}.

\begin{figure}[t]
 \includegraphics[height=0.47\textwidth, angle=270]{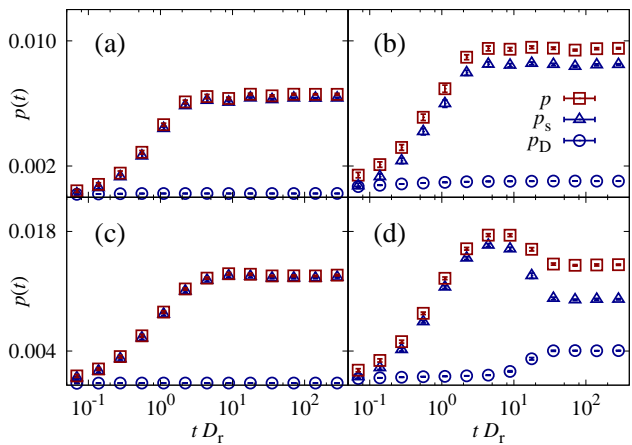}
\caption{Time evolution of the total pressure $p$ (red squares), swim pressure $p_\text{s}$ (blue triangles) and direct pressure $p_\text{D}$ (blue circles) for (a) $\ell_\text{p}/a = 25$ and $\phi = 0.2$, (b) $\ell_\text{p}/a = 25$ and $\phi = 0.35$, (c) $\ell_\text{p}/a = 50$ and $\phi = 0.2$, and (d) $\ell_\text{p}/a = 50$ and $\phi = 0.35$. Systems that remain homogeneous reach steady state on the timescale of $\tau_\text{r} = D_\text{r}^{-1}$ while phase-separating systems (in this case $\phi=0.35$, $\ell_\text{p}=50$, recall Fig.~\ref{fig:phase_diagram}) take orders of magnitude longer. Swim and total pressures grow non-montonically, first reaching a maximum value on the timescale of $\tau_\text{r}$, then falling to a lower steady-state value on a density-dependent timescale related to the increased variance of nucleation times at lower densities.
  \label{fig:p_vs_t_multi}} 
\end{figure}
\begin{figure*}[t]
\includegraphics[width=1\textwidth]{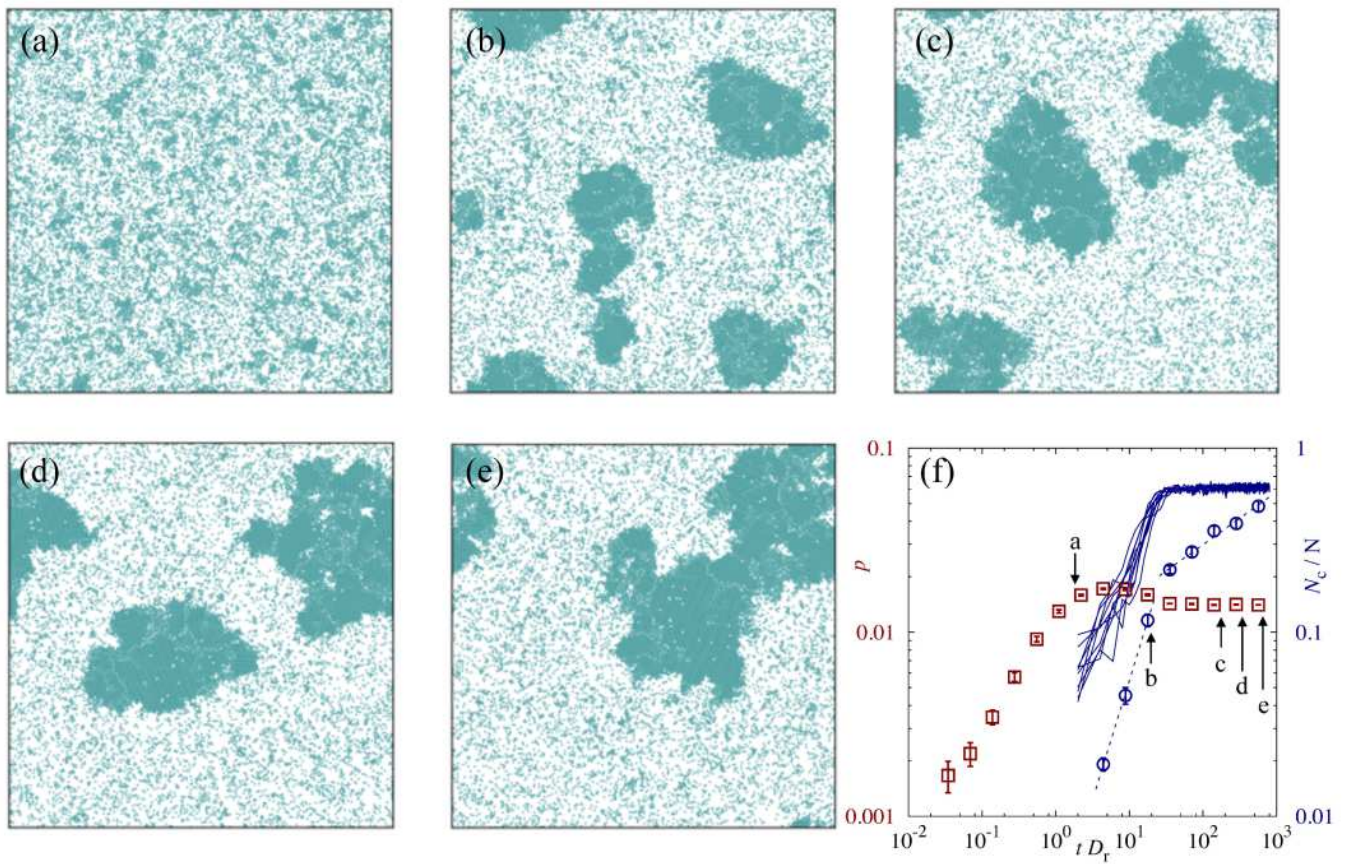}
\caption{Time evolution of pressure $p$ (red squares), total cluster fraction $N_\text{c}/N$ (blue lines), 
and largest cluster fraction $N_1/N$ (blue circles) for a system with size $L=400$, 
$\phi=0.35$, and $\ell_\text{p}=50$. $N_\text{c}/N$ is shown in (f) for 10 individual runs to emphasize its 
robust steady state. Snapshots (a--e) show a single run, for times that correspond
 to the points in the (f), labeled respectively. 
The overshoot in pressure results from the onset of clustering, which dampens the swim and total 
pressure in the long-time limit. The steady state of pressure corresponds with the steady state of $N_\text{c}$, while $N_1$ continues to grow as the system coarsens.
\label{fig:time_evolution_w_snapshots}} \end{figure*}

\section{Kinetics of Motility-Induced Phase separation}

The time evolution of the pressure from a random initial condition towards the final steady 
state reveals a non-monotonic dynamics directly related to the kinetics of MIPS.
This is displayed in Fig.~\ref{fig:p_vs_t_multi}. For parameter values corresponding to 
homogeneous steady states, the pressure evolves monotonically in time, reaching its 
steady state value in a time of the order of the persistence time, $\tau_\text{r}$. For 
parameter values corresponding to phase separated steady states, however, the convergence 
to the steady state is delayed by cluster nucleation and aggregation, which operate 
on a density-dependent timescale much longer than rotational diffusion. This results in 
the pressure temporarily overshooting its steady-state value, as shown in Fig.~\ref{fig:p_vs_t_multi}.  
The total time required for the pressure to reach steady state depends on density and, not 
surprisingly, is longest for the lowest densities that exhibit phase separation. The growth 
of the dense phase is associated with the formation of clusters of jammed particles. If 
the reorientation time $\tau_r$ required for particles to turn and escape from the cluster is 
longer than the mean free time between collisions, more particles will accumulate and the 
cluster will grow.  This process continues as long as the rate for escaping from the 
cluster ($D_\text{r}$), is slower than the rate at which particles absorb at the boundary, 
which is controlled by the collision rate. Systems with lower densities take a longer time to 
reach the final state due to lower nucleation and adsorption cross-sections. This is in agreement 
with recent work examining the kinetics of  MIPS in terms of classical nucleation theory~\cite{Redner2016}.

By comparing the evolution of pressure and of density correlations in our largest phase-separating systems ($L=400$), we can identify two distinct dynamical regimes. Once the 
pressure reaches its steady state value, clusters begin to coarsen at a slower rate,
corresponding to previous observations~\cite{Redner2013b,Stenhammar2013a}. 

By the time pressure equilibrates, multiple clusters have formed and there is an average zero 
net flux between phases, shown in Fig.~\ref{fig:time_evolution_w_snapshots} where we additionally
quantify the number of particles in the largest cluster $N_1$ alongside all clustered particles $N_\text{c}$. 
At this stage, the dense aggregates move very slowly, having caged the motility of most aggregated
particles, only allowing those at the boundary to de-adsorb, while large clusters coalesce into 
one another until system-spanning phase separation occurs. We find a division in time regimes 
of growth for $N_\text{c}(t)$ and $N_1 (t)$ and show snapshots of this process in
Fig.~\ref{fig:time_evolution_w_snapshots}. The largest cluster continues to grow long after the 
cluster fraction saturates, which corresponds with pressure reaching its steady state.

In particular, for the data shown in Fig.~\ref{fig:time_evolution_w_snapshots}, we have fit 
the number of particles in the largest cluster to
\begin{equation}
N_1(t) \sim t^\gamma.
\end{equation}
In the first time regime (up to and including point B in the figure, that is, for $t D_\text{r}< 20$) this fit gives
$\gamma_1=1.3(2), \chi^2/\text{d.o.f.}=0.12/1$. In the coarsening regime, for $tD_\text{r}> 20$, we obtain $\gamma_2=0.29(4), \chi^2/\text{d.o.f.}=1.65/3$.
These fits are plotted with dotted lines in the figure.

We note that the coarsening regime has been studied in previous work~\cite{Redner2013b,Stenhammar2013a} using the growth of a length scale computed from the structure factor: $L(t)\propto t^\alpha$, $\alpha\approx0.28$. If we reproduce the analysis 
of $L(t)$ in~\cite{Stenhammar2013a} with our data we obtain $\alpha = 0.26(2), \chi^2/\text{d.o.f.}=1.82/3$ for $tD_\text{r}>20$.

\begin{figure}[]
    \includegraphics[height=0.48\textwidth, angle=270]{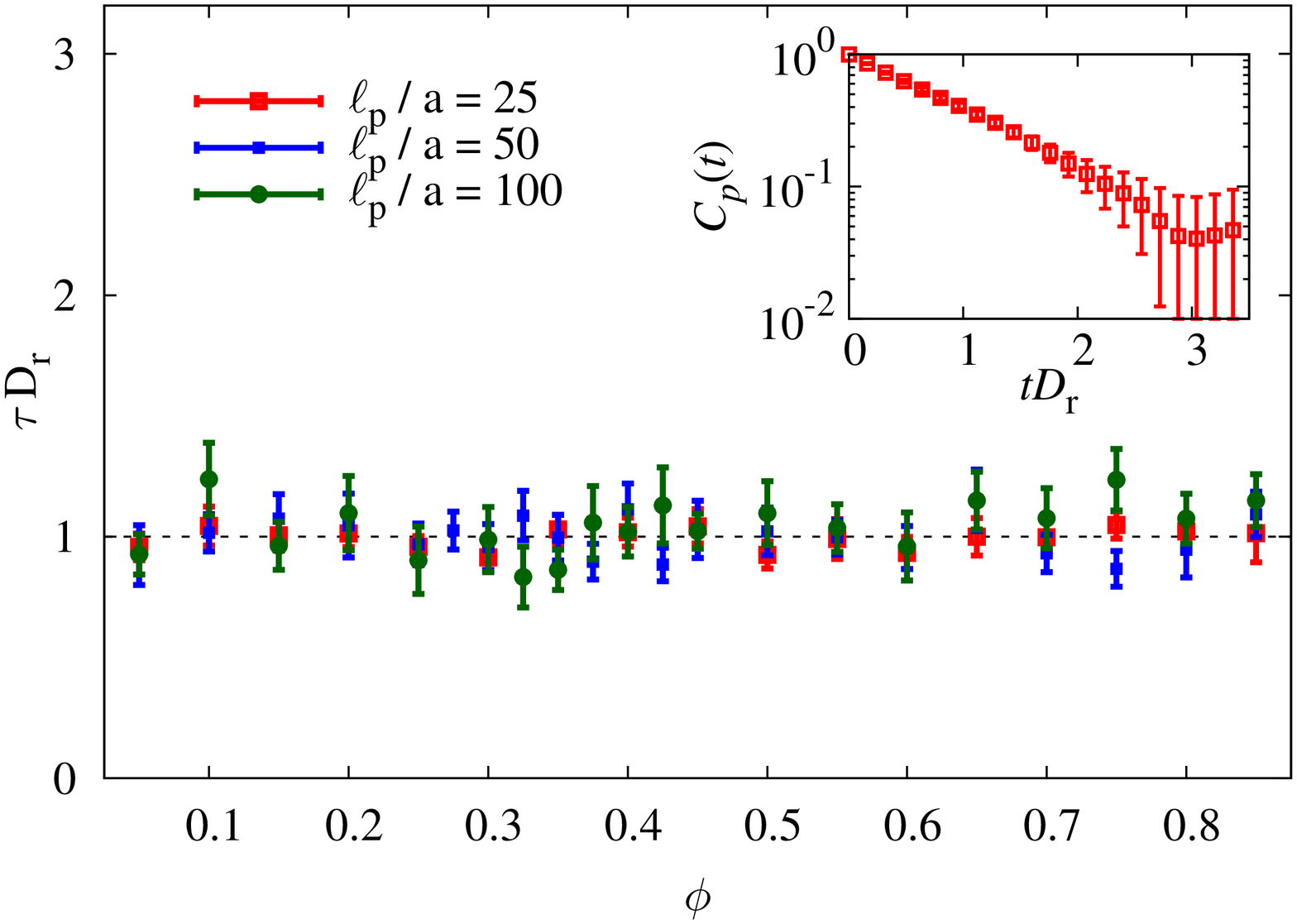}
\caption{Correlation time $\tau$ for $\ell_\text{p}=25,50,100$
computed for the correlation function, Eq.~\eqref{eq:corr}. 
Inset shows an example of the steady-state correlation function, for $\phi=0.4$ and $\ell_\text{p} = 25$.
Even though the time needed to reach the steady state depends on $\phi$
(Figure~\ref{fig:p_vs_t_multi}), once the steady-state has been reached
$\tau$ is equal to the single particle persistence time,
$\tau=\tau_\text{r}= \ell_\text{p}/v_0$.
\label{fig:tau_corr}}
\end{figure}

Finally, we compare the approach to the steady state to the dynamics  of pressure fluctuations 
in the steady state by computing the time autocorrelation function of the instantaneous mean 
pressure, given by

\begin{align}\label{eq:corr}
C_p(t) =\frac{ \langle p(s) p(s+t) \rangle - \langle p\rangle^2}{\langle p^2 \rangle - \langle p\rangle^2}\, .
\end{align} 

This function is shown in Fig.~\ref{fig:tau_corr} for $\ell_p=25$ and $\phi=0.4$. The
correlation function decays exponentially, allowing us to extract a relaxation
time $\tau$, \begin{equation} C_p(t) \simeq \text{e}^{-t/\tau}\;,
\end{equation}

We find that $\tau$ does not depend on density and coincides with $\tau_r$ for
all our runs (Fig.~\ref{fig:tau_corr}). Therefore, even though the time evolution of the 
pressure in the approach to the steady state is density dependent, once the steady state 
has been reached the only time scale for pressure fluctuations is given by the single-particle 
rotational diffusion.

\section{Conclusions}

We have examined the effects of finite system size and of the  kinetics of MIPS on the pressure of ABPs in two dimensions. In a dilute gas of ABPs the finite-size effects on pressure for both 
open (periodic) and closed (particles in a square box) boundary conditions are quantitatively similar to what one would expect for a thermal gas. At finite density, 
finite-size effects are pronounced even for the periodic case. We 
find that the box size has to be several times larger than the persistence length 
of the particles to obtain results representative of bulk behavior. 
This has implications for studies in strip geometries~\cite{Bialke2015}, where a careful control is needed to avoid spurious anisotropies in the stresses.

For parameter values corresponding to MIPS, we have examined the correlation between the relaxation of the mean pressure to its steady state value and the kinetics of  
clustering and phase separation. In this regime, the interplay between
the decreasing swim pressure  and the increasing direct pressure from interaction in the incipient clusters results
in long, density-dependent time scales for approach to the steady state. The phase separation process shows two distinct dynamical regimes: a rapid growth corresponding to the formation of small clusters, followed by a slower regime of cluster coalescence and coarsening. These two regimes are reflected in the time evolution of the pressure that first grows rapidly during the small cluster nucleation when it is mainly controlled by the swim pressure of the gas, even overshooting its steady state value, and then remains constant during coarsening when the net flux of particles between the 
dense and dilute phases vanishes. The overshoot of the swim pressure upon phase separation is a distinctive feature of active particles associated with crowding of the gas at the boundary of the dense phase. The cluster provides an effective bounding wall to the active gas, strongly suppressing the swim pressure.  This effect has no analogue in thermal systems where the kinetic contribution to the pressure is not affected by interactions nor by interfaces.

The total pressure, has, however, been shown to remain equal across the two phases, confirming that this minimal model can be described in terms of an effective thermodynamics. The relationship between the kinetics of MIPS and the relaxation of pressure to its steady state value additionally validates the use of equilibrium-like ideas as done in Ref.~\cite{Redner2016} to describe the coarsening kinetics.

NSF-DMR-1305184, NSF-DGE-1068780, ACI-1341006, FIS2015-65078-C02, BIFI-ZCAM

MCM was supported by NSF-DMR-1305184. MCM and AP acknowledge support by the NSF
IGERT program through award  NSF-DGE-1068780. MCM, AP and DY  were additionally
supported by the Soft Matter Program at Syracuse University. Our simulations
were carried out on the Syracuse University HTC Campus Grid which is supported
by NSF award ACI-1341006 and on the Memento supercomputer.  DY acknowledges support 
 by the Ministerio de Ciencia y Tecnolog\'ia (Spain) through grant no.
FIS2015-65078-C02 as well as the resources, technical expertise and assistance provided by
BIFI-ZCAM (Universidad de Zaragoza).

\bibliographystyle{apsrev4-1}
%

\end{document}